\documentclass{elsart}
 
\begin{document}
 \begin{frontmatter}
 \input epsf
 \title{\large{Self-Organized Criticality in a Fibre-Bundle type
model}}
 \author[FT]{Y. Moreno\thanksref{alsoat}},
 \author[CT]{ J.B.G\'{o}mez},
 \author[FT]{A. F. Pacheco}
 \address[FT]{Departamento de F\'{\i}sica Te\'{o}rica, Universidad de Zaragoza, 50009
Zaragoza, Spain.}
 \address[CT]{ Departamento de Ciencias de la Tierra,
 Universidad de Zaragoza, 50009 Zaragoza, Spain.}
 \thanks[alsoat]{On leave from Departamento de F\'{\i}sica, Technological
University of Havana, ISPJAE, Havana 19390, Cuba.}
\begin{abstract} The dynamics of a fibre-bundle type model with equal load sharing rule
is numerically studied. The system, formed by $N$ elements, is driven by a slow increase
of the load upon it which is removed in a novel way through internal transfers to the
elements broken during avalanches. When an avalanche ends, failed elements are
regenerated with strengths taken from a probability distribution. For a large enough $N$
and certain restrictions on the distribution of individual strengths, the system reaches
a self-organized critical state where the spectrum of avalanche sizes is a power law with
an exponent $\tau\simeq 1.5$.
 \end{abstract}
\end{frontmatter}

\section{Introduction}
\label{intro}

Twelve years ago the idea of self-organized criticality (SOC) was introduced by Bak, Tang
and Wiesenfeld \cite{bak87}, as a way of understanding the fractal structure and the
$1/f$ noise behaviour displayed by a wide variety of large interactive systems. Although
a precise definition of SOC is still lacking, many papers on the subject have appeared
\cite{jensen98} and avalanche-like behaviour has been experimentally observed in many
real physical phenomena: microfracturing processes \cite{zapperi97}, earthquakes
\cite{gutenberg56}, fluid flow through porous media \cite{sahimi93}, flux lines in
superconductors \cite{field95}, etc. Among the number of proposed models to describe SOC
behaviour, the sandpile \cite{bak87}, forest-fire \cite{clar96}, invasion percolation
\cite{wilkinson83}, Bak-Sneppen \cite{bak93} and Olami, Feder and Christensen (OFC)
\cite{olami92} type models have been perhaps the most intensively studied, constituting
paradigms in this subject.

In the present paper we make use of the well-known fibre-bundle models widely used to
analyze the fracture process in heterogeneous materials \cite{herrmann90}. They have also
been applied in geophysics \cite{newman94,turcotte97}. In fibre-bundle models, a set of
elements is located on a supporting lattice, each with a strength threshold sampled from
a probability distribution. In these models, once an element fails, its load is
distributed among the surviving elements. Different load transfer rules can be defined
depending on the range of the interaction assumed. In the ELS (for equal load sharing)
case, the load carried by a failed element is equally distributed among the surviving
elements of the system, representing in this way a long-range interaction. The ELS model
is a sort of mean field approximation to the more realistic local transfer schemes. It
has been found that the distribution of avalanche sizes in a breaking cycle in the static
version of the ELS model follows a power law \cite{hemmer92}. It is clear that this is
not at all a model of the SOC type, because in the breaking process a stationary state
can not be reached as the broken elements remain broken during the cycle. In what follows
we propose a fibre-bundle model which does exhibit SOC behaviour by using an ELS transfer
rule, a novel way for dissipation, and the hypothesis that the failed elements after an
avalanche are regenerated, i.e., they are assigned new strength thresholds. In Section 2
we present the model. Section 3 is devoted to present the results obtained from
simulations and to discussion. Our conclusions are given in Section 4.

\section{The Model}
\label{model}

Let there be a set of $N$ elements located on a supporting lattice.
Suppose that each element carries a given load $\sigma$ and has a
strength threshold $\sigma_{th}$. This can be viewed as a
representation of a disordered material in which each small volume is
described by its breaking characteristics. In order to assign the
random thresholds, different probability distributions can be
considered. In materials science the Weibull distribution is usually
used,
\[
P(\sigma)=1-e^{-(\frac{\sigma}{\sigma_{0}})^{\rho}},
\]
$\rho$ being the so-called Weibull index, which controls the degree of
threshold disorder in the system (the bigger the Weibull index, the
narrower the range of threshold values), and $\sigma_{0}$ is a load of
reference which acts as unity. In the following we will assume
$\sigma_{0}=1$, and therefore the loads and thresholds used henceforth
are dimensionless. Thus, to each site $i$, $1\leq i\leq N$, one
assigns a random threshold value $\sigma_{i_{th}}$:
\begin{equation}
n_i=1-e^{-\sigma_{i_{th}}^{\rho}},
\end{equation}
where $n_i$ are random numbers uniformly distributed between $0$ and
$1$.

At the beginning, the load carried by all the elements is set to zero.
In analogy with the OFC model \cite{olami92} the system is driven at
the same rate. During each {\it external} time step of loading, all
the elements in the set increase their load by a small amount $\nu$,
\begin{equation}
\sigma_{i}\rightarrow\sigma_{i}+\nu\quad,\quad\forall i.
\label{two}
\end{equation}
This mode of driving the system allows us to obtain the limit of
infinitesimal driving rate. In practice, we search for the smallest
threshold value and add this amount to all the elements of the system.
This makes, at least, one element critical. Suppose that as a
consequence of applying (\ref{two}), $q_{1}$ elements become unstable
(usually, $q_1=1$). The homogeneous drive is switched off, the
unstable elements fail and the following relaxing rule is applied to
all the $q_{1}$ elements:
\[
\sigma_{i}\rightarrow 0\quad,\quad\forall\sigma_{i}\geq\sigma_{i_{th}}.
\]
Now, assuming an ELS transfer rule, the total load supported by the
$q_{1}$ elements, $\sigma_{dist}=\sum\limits_{i=1}^{q_{1}}\sigma_{i}$,
is equally distributed among all the remaining elements (the $N-q_{1}$
surviving elements), so that the new load on all the surviving
elements is
\[
\sigma_{i}\rightarrow\sigma_{i}+\frac{\sigma_{dist}}{N-q_{1}}\quad,\quad\forall\sigma_{i}<\sigma_{i_{th}}.
\]
This may have the effect that other elements become unstable and the
avalanche continues. This case will be commented on in the next
paragraph. If this is not the case, the broken elements are
regenerated with new random threshold values and zero load, and rule\
(\ref{two}) is repeated until a new avalanche is triggered. We define
the size, $s$, of an avalanche as the number of broken elements
between two successive steps of external loading of the system, an
{\it internal} time step as a visit to all the $N$ elements of the set
checking whether or not their $\sigma$-value is larger than or equal
to their $\sigma_{th}$-value, and the avalanche lifetime, $T$, as the
number of internal time steps needed for the system to be completely
relaxed.

Now let us assume that as a consequence of the distribution of the
amount $\sigma_{dist}$, $q_2$ elements became overcritical, i.e,
$\sigma_{i}\geq\sigma_{i_{th}}$, for these $q_2$ elements. Being part
of the same avalanche, the $q_1$ elements broken before are not
regenerated and the new amount of load to be distributed is
$\sigma_{dist}=\sum\limits_{i=1}^{q_2}\sigma_{i}$. As mentioned
before, the ELS transfer scheme implies that the load supported by
failing elements is equally distributed among the surviving elements
of the set. On the other hand, it is clear that to make possible the
existence of a stationary state one has to introduce an exit of load
to make it possible that, on average, the load inflow is compensated
by an outflow from the system. We will assume that the system loses
load through the elements that have previously failed in the {\it
same} avalanche, that is, when the transfer of the load carried by
currently failing elements takes place, the portion of load that
corresponds to the already broken elements in previous internal time
steps of the same avalanche leaves the system. This is a novel way for
dissipation and plays the role of the boundaries in other models of
SOC.  Its physical meaning is straightforward: regions that have just
failed in that avalanche can not accumulate stress during the same
fracture process. This assumption implies that the total amount of
load removed from the system after an avalanche has ended depends on
both the avalanche size and on its lifetime.

Continuing the process, the amount $\sigma_{dist}$ is distributed
among all the $N-q_2$ elements which remain as spectators in this
second internal time step of the breaking process, that is, the $q_1$
elements broken in the first internal time step, and the remaining
$N-q_{2}-q_{1}$ elements which are stable. Hence, the update for the
surviving elements is performed according to:
\begin{equation}
\sigma_{i}\rightarrow\sigma_{i}+\frac{\sigma_{dist}}{N-q_{2}}\quad,\quad\forall\sigma_{i}<\sigma_{i_{th}}
\label{three}
\end{equation}
and the load $\sigma_{lost}=q_{1}\frac{\sigma_{dist}}{(N-q_{2})}$,
corresponding to the $q_{1}$ broken elements failed in the first
internal time step, is lost. The surviving elements are checked again.
If, for example, $q_{3}$ new elements become unstable in this third
internal time step,
$\frac{1}{(N-q_{3})}\sum\limits_{i}^{q_{3}}\sigma_{i}$ units of load
are added to the remaining $N-q_{3}-q_{2}-q_{1}$ surviving elements
and
$\sigma_{lost}=\frac{(q_2+q_1)}{N-q_3}\sum\limits_{i=1}^{q_3}\sigma_{i}$
units of load are lost in this third internal time step. We check if
new elements become unstable and so on. The process continues until we
regain a static state ($\sigma_{i}<\sigma_{i_{th}}$ for all the
surviving elements) where the avalanche ends. The broken elements are
regenerated with new randomly chosen strength values and with loads
equal to zero. In the example given above, if no elements become
unstable when the third distribution of load takes place, the
avalanche stops and its size and lifetime are $s=q_1+q_2+q_3$ and
$T=3$, respectively.

\section{Results and Discussion}
\label{results}

Two distinct behaviours of the system explained in Section 2 are
obtained from numerical simulations according to the width of the
probability distribution from which the strength thresholds are
taken. In the first, the system does not exhibit the
characteristics of SOC behaviour although the avalanche size
distribution for small avalanches is a power law, while in the
second one the system is able to settle into a stationary state
with fluctuations around the temporal mean value of the system
load, and with power-law distributions for both the avalanche
sizes (over the entire range of avalanche sizes) and the avalanche
lifetimes.

\subsection{Non-SOC behaviour} \label{nonsoc}

For large values of $\rho$, the Weibull distribution is sharply peaked. Thus,
irrespective of the system size, there are many elements in the set with similar breaking
properties, i.e., with very close strength threshold values. With the hypotheses of our
model, it provokes a simple pattern of dynamical evolution: periods of slow loading
followed by catastrophic avalanches. This is observed in Fig.\ \ref{figure1}, where we
have plotted the value of the mean load per element stored in the system as a function of
the number of avalanches for a system of $N=10000$ elements and $\rho=4$. As can be
observed, the average value of the system load does not reach a statistically stationary
value but systematically increases with time and suddenly falls to zero, with a sort of
quasi-periodic sawtooth behaviour (the pattern is not completely periodic, because there
is some fluctuation in the amplitude of the drop and in the time interval between major
avalanches).

We have also monitored the distribution of avalanche sizes. The
results obtained are similar to those reported in
Ref.\cite{knopoff92}, where the dynamics of a sandpile-like model was
investigated, and with those of Ref.\cite{hemmer92}. Fig.\
\ref{figure2} shows, in dimensionless units, the avalanche size
distribution for system sizes of $N=50, 100, 1000$ and $10000$ with
$\rho=4$. Two distinct features are observed. For the smallest system
size there are avalanches of almost all sizes and the distribution has
a sharp peak for large values of avalanche sizes. Increasing the
system size produces a gap in the avalanche spectrum. Now, the event
size distribution is bimodal: at the small scale, the spectrum of
small avalanches is of the power law type over a reduced range of
avalanche sizes, whereas at the largest scale there is an excess of
events whose sizes are of the order of the system size.

The absence of avalanches for intermediate sizes is clear and can be interpreted as being
due to the simultaneous failure of many elements with very close threshold values whose
failure triggers catastrophic events of large sizes. It is interesting to note that
whenever a power law distribution can be identified, its exponent is of about
$\frac{5}{2}$, i.e., the same reported from an analytic study for the burst distribution
in static fibre-bundle models with ELS transfer rule \cite{hemmer92}. This is
understandable because the behaviour of our model in this non-SOC regime is similar to a
succession of breakings of static ELS fibre-bundle sets. There \cite{hemmer92}, at the
beginning, small avalanches are produced randomly dispersed throughout the system; then a
crack is nucleated that leads to a final, catastrophic avalanche where an important
fraction of $N$ fails at the same instant. As our system dissipates through the elements
broken during the ongoing avalanche, this final catastrophic avalanche unloads the system
very efficiently, leading it to the beginning of a new cycle of slow loading. This
process (Fig.\ \ref{figure1}) could be called the ``cistern effect", because of its
similarity with the familiar rate of filling and flushing of an old-fashioned toilet
cistern.

\subsection{SOC behaviour} \label{soc}

For $\rho$ values such that the width of the Weibull distribution is wider, the system,
without any tuning, is able to self-organize into a stationary state where the flow of
load into the system equals the flow of load out of the system. This is due to the large
inhomogenities in the distribution of threshold values, that is, the existence of both
highly resistant elements together with other very brittle elements. Fig.\ \ref{figure3}
shows the evolution of the load per element accumulated by the whole system as a function
of the number of avalanches for systems of $N=10000$ and $N=50000$ elements with $\rho=2$
in both cases. As can be seen, the average value of the stored load fluctuates around a
temporal mean value and these fluctuations decrease as the size of the system increases.
This stationary state is reached if the size of the system is large enough. For a too
small  $N$ there exists a background of avalanches, with sizes of the order of the system
size, which frequently provokes total collapses. This fact agrees with the basic
assumption that SOC behaviour demands large systems.

In Fig.\ \ref{figure4} we have plotted the avalanche size distribution
for a system with $N=50000$ elements and $\rho=2$. A power law of the
form $P(s)\sim s^{-\tau}$ can be very well fitted over the entire
range of event sizes with a critical exponent $\tau$ close to $1.5$.
We have also found a power law $P(T)\sim T^{-y}$ for the distribution
of avalanche lifetimes $T$, shown in Fig.\ \ref{figure5} for values as
in Fig.\ \ref{figure4}. Table 1. summarizes the results obtained for
different values of $\rho$ in a system of $N=50000$ elements.

We have  checked that once the system fulfills  the SOC behaviour (appropiate $\rho$, and
large enough $N$), the critical exponents obtained of both the avalanche size
distribution and the avalanche lifetime distribution do not vary with the system size.
The exponent $y$ characterizing the avalanche lifetime distribution slightly varies with
the Weibull index. However, the value of the critical exponent $\tau$ for the
distribution of avalanche sizes is universal, that is, it does not depend on the $\rho$
value. The result $\tau\sim1.5$ is close to the value derived from mean field
approximations for SOC systems \cite{vespignani98}. This fact is not surprising because
as mentioned above the ELS transfer scheme is itself a sort of mean field-like
approximation.

We have performed numerical simulations of this model, changing the
Weibull probability distribution by a power-law (Pareto-like)
distribution,
\[
P(\sigma)=1-\frac{1}{(\frac{\sigma}{\sigma_{0}})^{m}+1},
\]
where $m$ and $\sigma_{0}$ are the parameters of the distribution. In
this case $m$ plays the role of $\rho$ in the Weibull distribution and
$\sigma_{0}$ is again a load of reference which is set to one in the
numerical simulations. Our results appear in Table 2. The dynamical
evolution of the system is qualitatively similar and the avalanche
size distribution shows a power law with, again, a critical exponent
$\tau\simeq 1.5$.

\section{CONCLUSIONS}
\label{Conclu}

We have introduced  a fibre-bundle type model with an ELS transfer rule and a new way for
load removal from the system. The qualitative results obtained for the dynamics of the
model are summarized in a type of two-phase diagram as can be seen in Fig.\
\ref{figure6}.

For small values of $N$, no matter how the threshold values are distributed, the dynamics
of the system is of the type of the static fibre-bundle model with ELS transfer rule.
Increasing the value of $\rho$, even for large $N$ the system is unable to avoid a
quasiperiodic sequence of complete failure. This non-SOC zone is characterized by an
avalanche size distribution with two clear features: one for the smaller scales in which
the avalanche size distribution is of the power-law type with a critical exponent
$\tau=2.5$ and the other for the bigger scales with a sharply peaked distribution in the
neigbourhood of $N$.

On the other hand, for large enough $N$ and in the range of moderate
$\rho$-values corresponding to large inhomogeneities in the threshold
values of the elements, the system self-organizes into a statistically
stationary state characterized by power law distributions for the size
and duration of the avalanches. We have found that the critical
exponent characterizing the avalanche size distribution is
$\tau\sim1.5$, very close to the value derived from mean field
arguments for SOC systems, and that it is universal, that is, it does
not depend on the probability distribution from which the threshold
values are taken. Finally, we shall say that for large values of
$\rho$ and $N$ the exploration of the dynamics of the system becomes
very difficult because of the very long transient period needed before
reliable conclusions can be obtained.

\section{ACKNOWLEDGMENTS}

We thanks M. V\'{a}zquez-Prada for discussions. Y. M. thanks the AECI
for financial support. This work was supported in part by the Spanish
DGICYT.

\newpage

\begin{figure}
\caption{Dimensionless mean load per element stored in the surviving elements after an
avalanche ends, as a function of the number of avalanches ($\rho=4$). Note the
quasi-periodic sawtooth behaviour. This is a typical graph for the non-SOC state.}
\label{figure1}
\end{figure}

\newpage

\begin{figure}
\caption{Avalanche size distributions for different system sizes ($\rho=4$). The solid line has a slope of $-2.5$.}
\label{figure2}
\end{figure}

\newpage

\begin{figure}
\caption{Mean load per element carried by surviving elements when an avalache ends
as a function of the number of avalanches. Fluctuations around an
average value (represented by the horizontal solid line) decrease as
the size of the system increases from $N=10000$ elements (dotted line)
to $N=50000$ elements (solid line). $\rho=2$ in both cases.}
\label{figure3}
\end{figure}

\newpage

\begin{figure}
\caption{Typical graph of the avalanche size distribution in the SOC regime. This case
is for a system of $N=50000$ elements and $\rho=2$. The straight line
has a slope $-\tau=-1.5$.}
\label{figure4}
\end{figure}

\newpage

\begin{figure}
\caption{Avalanche lifetime distribution for a system consisting of $N=50000$ elements
and $\rho=2$. The slope of the solid line is $-y=-1.81$.}
\label{figure5}
\end{figure}

\newpage

\begin{figure}
\caption{Schematic phase diagram of the system. Inset graphs summarize typical avalanche
size distribution and load fluctuations for the two regimes.}
\label{figure6}
\end{figure}

 \vspace{15cm}



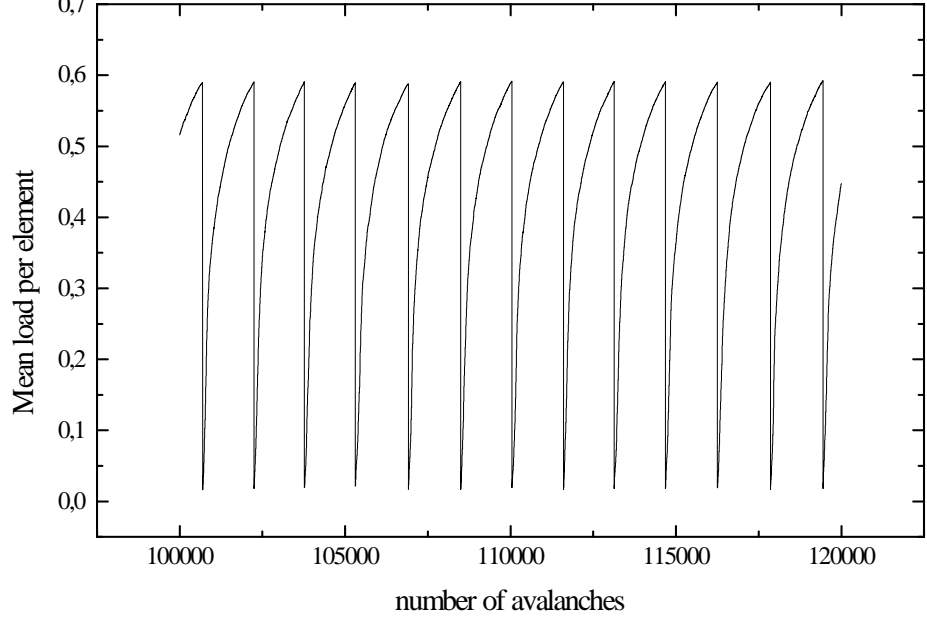
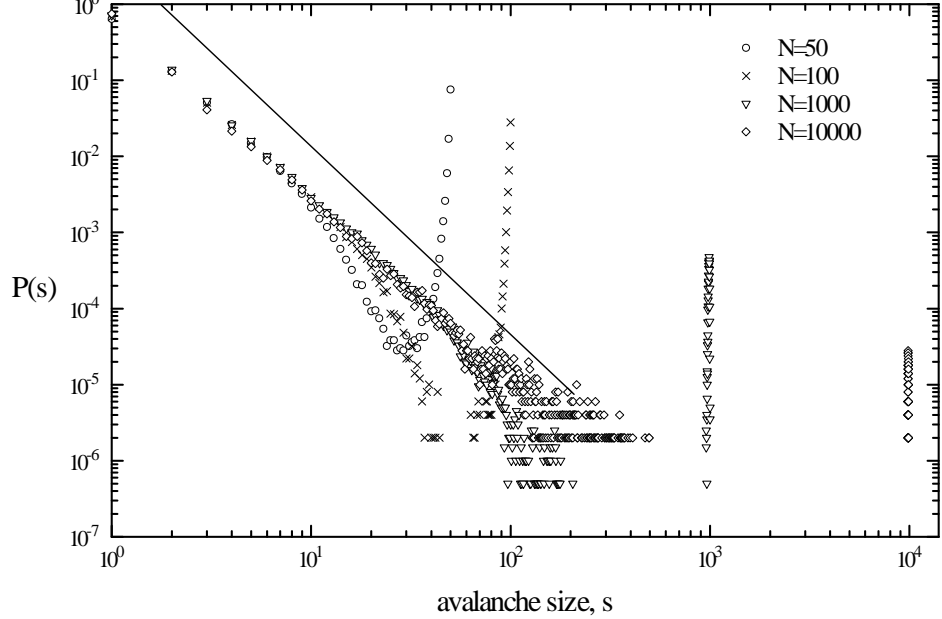
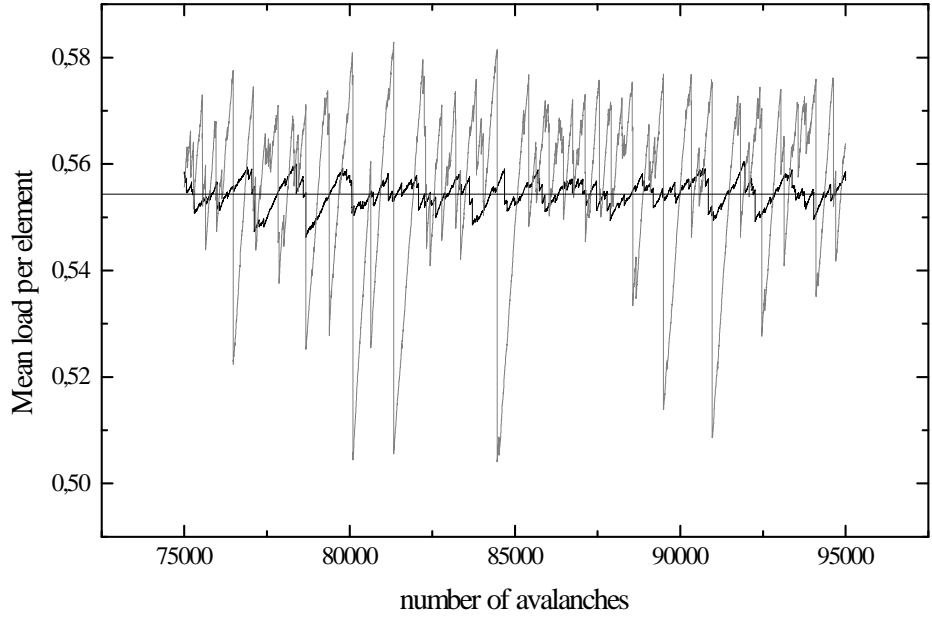
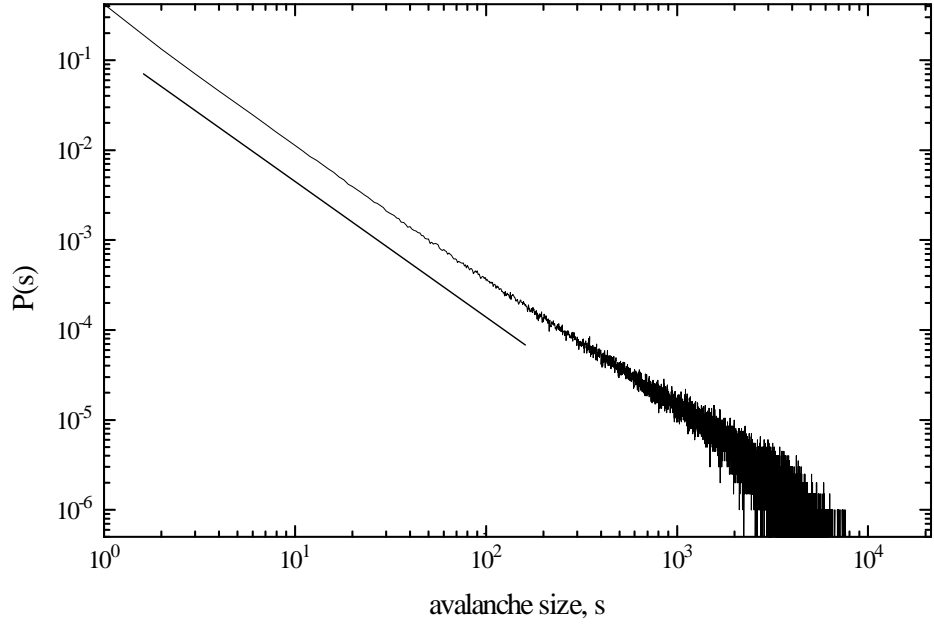
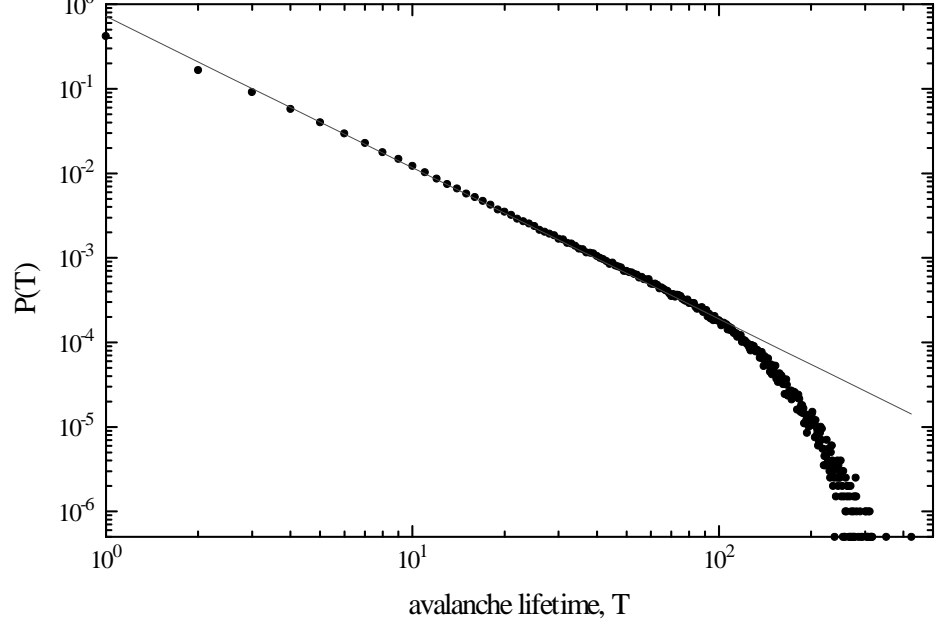
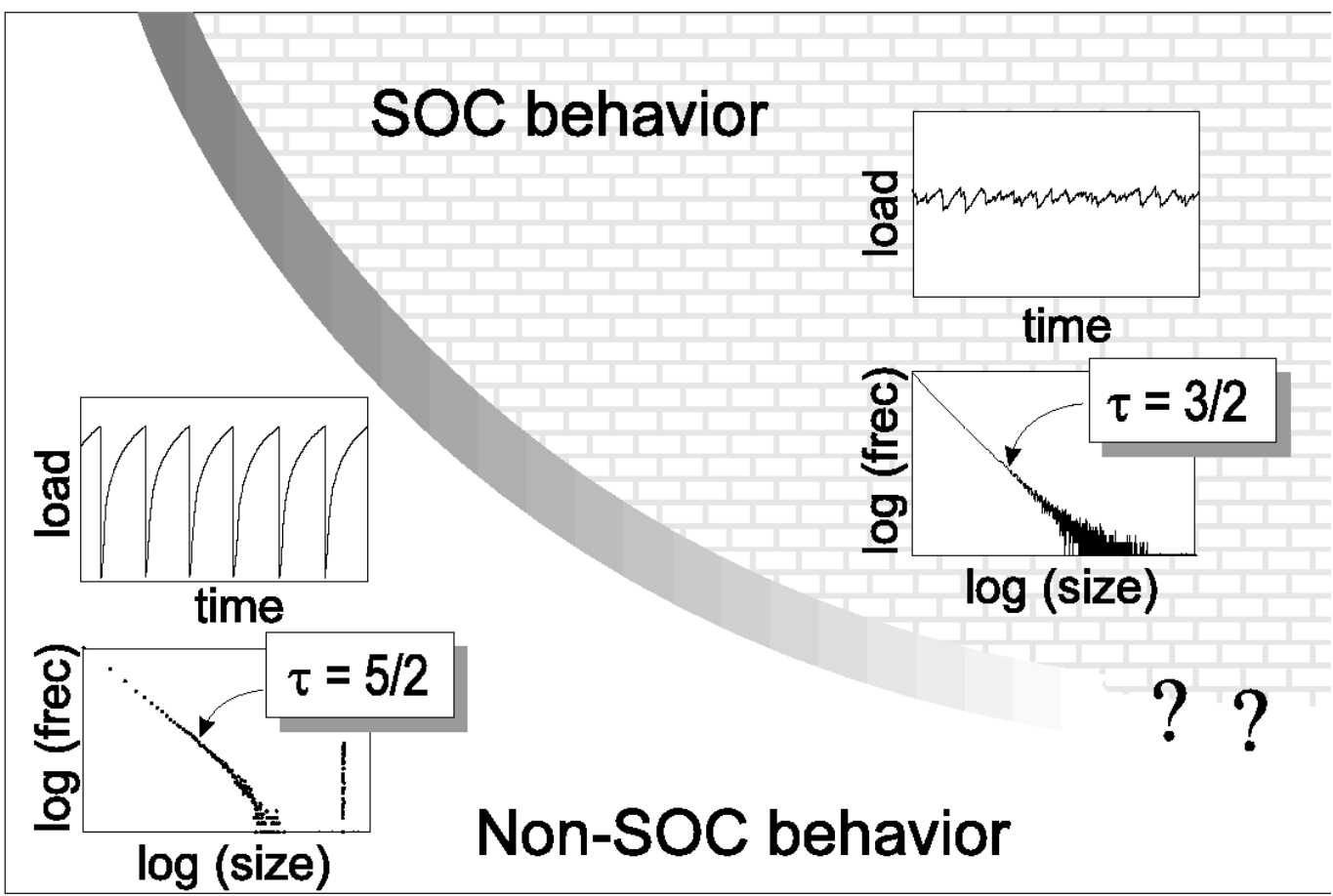
\end{document}